 \magnification=\magstep1 
 \font\romsix=cmr6 scaled\magstep0
 \font\romnine=cmr9 scaled\magstep0
 
 \font\medbigrom=cmr10 scaled\magstep1
 \def\romtwelve{\medbigrom}
 \def\frac#1#2{{#1 \over #2}}
 \def\standardpage{\hsize=6 true in \vsize= 8.5 true in \hoffset=0.25 true in}
 \def\tenstart{\normalbase \tenrm ~~~~~~} 
 \overfullrule=0pt  \def\refmark{}  \def\eq#1{Eq.\ (#1)} 
 \def\NI{\noindent}  \def\pritem#1{\vskip .04in\item{[#1]}}
 \def\upref#1{$^{#1}$} 
 \newcount\ftnum  \newcount\checknum  \newif\ifeqnerr
 \def\aq#1{\global\advance\ftnum by 1 \xdef#1{\the\ftnum}}
 \def\neweq#1{{\global\advance\checknum by 1
    \edef\chqtmp{\the\checknum}
    \checkeq#1}}
\def\checkeq#1{\ifx#1\chqtmp\global\eqnerrfalse
    \else \global\eqnerrtrue
        \message{ALLOCATION ERROR for \string#1:
             preassigned #1, in sequence \chqtmp.}\fi}

\def\aqx#1#2{\xdef#2{\the\ftnum#1}}
\def\neweqx#1#2{{\edef\chqtmp{\the\checknum#1}
    \checkeq#2}}

   \def\normalbase{\baselineskip 16pt}\standardpage 
   \def\bsheader#1{{
   \removelastskip\vskip 15pt plus 20pt \penalty-200 \vskip 0pt plus -16pt
   \NI\bf #1}\nobreak\medskip\nobreak}
   \footline={\ifnum\pageno=1 \hfill \else\hss\tenrm\folio\hss\fi}
   \def\mathname#1{``$#1$"} \def\M{{\cal M}}
   \def\Psitest{\Psi_{\hbox{\romsix test}}} \def\NN{{\cal N}}
   \def\lmax{\lambda_{\hbox{\romsix max}}}
   \tenstart   
  
   \aq\chuang \aq\timebook \aq\destruct \aq\measure \aq\error \aq\primitive
  \aq\NOg 
  \aq\yanase \aq\pibook \aq\bigdeal \aq\notmerely \aq\grabert
  
 \centerline{\romtwelve
               BOUNDS ON DECOHERENCE AND ERROR }
 \bigskip \centerline {L. S. Schulman}
 \smallskip
 \centerline {\romnine
                 Physics Department, Clarkson University }
 \centerline {\romnine
                 Potsdam, NY 13699-5820 USA }
 \medskip
  
 {\narrower \romnine \baselineskip 10pt
 \centerline{ABSTRACT} \vskip 2pt
 When a confined system interacts with its walls (treated quantum
mechanically), there is an intertwining of degrees of freedom. We show that
{\it this need not lead to entanglement}, hence decoherence. It {\it will\/}
generally lead to error. The wave function optimization required to avoid
decoherence is also examined.
 \par} \medskip

\NI

 \bsheader{Introduction}

   Physical implementation of quantum computing
algorithms,\upref{\chuang\refmark} experimental tests of certain
theories,\upref{\timebook\refmark} as well as other contemporary problems,
require that for fairly large systems the time evolution be fully described by
$\psi \to \exp(-iHt/\hbar)\psi$, with no ``measurement," or to be more precise,
no decoherence or interaction with the environment. Such interaction can cause
entanglement with environmental degrees of freedom and prevent interference
with portions of the wave function that have not experienced the identical
interaction. Moreover, those same interactions can induce errors, that is,
system wave function different from that providing the desired output,
computational or otherwise.

  For any laboratory system one can expect a degree of entanglement with the
environment, simply due to the fact that the system is pinned to the table.  In
particular, when part of the system rebounds from the walls confining it (even
electromagnetic walls) conservation of momentum demands an intertwining of the
degrees of freedom.

   Taking the approach in [\destruct\refmark], I here begin from this
inevitable intertwining and establish the extent to which it leads to
entanglement. The measure of entanglement is that given in [\measure\refmark].
There is a surprise in the calculation: for appropriately tailored wave
functions, there need be no decoherence! This leads us to explore the
significance of the tailoring. However, although {\it decoherence} is
avoidable, we will show that {\it error} is not.\upref{\error\refmark}

  Whether the decoherence is large or small (for nearly matching wave functions
it is of order system/container mass ratio) the resulting amplitude defect must
be subtracted from the wave function for {\it each\/} collision, allowing for
physically significant effects.

 \bsheader{Interacting with a wall}

   A confined system will, from time to time, interact with its container.
Dissipative walls, in the sense that the interaction is an inelastic collision,
immediately lead to significant entanglement; for our bounds we therefore
assume that the collision is elastic and involves no degree of freedom beyond
that required to contain the system. Our model is therefore the scattering of
two point particles, one small (mass $m$) representing a piece of the quantum
computer, and one large (mass $M$), representing the
container.\upref{\primitive\refmark}
 
   Before the collision we assume the wave function to be unentangled, that is,
$\Psi_I = \Gamma(X)\Phi(x)$, with position variables $X$ and $x$ corresponding
to the large and small masses, respectively. We make several simplifying
assumptions: 1.)~~Restriction to one dimension, reasonable if the large
``particle" is in fact a wall. 2.)~~Rapid completion of the scattering.
3)~~Short range, infinite repulsion. 4.)~~Gaussian wave packets.  Assumptions
\#2 and \#3 would not be true in detail, but I expect that departures from them
will only make things {\it worse} (with respect to decoherence and error).  We
further comment below on these assumptions. 

   If the interaction with the wall could be treated as a pure
potential-interaction with a fixed object, the wave function after the
collision would be\upref{\NOg\refmark} $\Gamma(X)\Phi(-x)$. On the other hand,
the correct form of the final wave function can be seen by going to center of
mass coordinates, $R=(MX+mx)/\M$, $u=x-X$, with $\M=M+m$. In these coordinates
$\Psi_I = \Gamma(R-\delta u)\Phi(R+\gamma u)$, where $\delta=m/\M$ and
$\gamma=M/\M$. With the above assumptions, the wave function {\it after\/} the
collision is $\Psi_F = \Gamma(R+\delta u)\Phi(R-\gamma u)$, i.e., $u\to-u$. To
show this, recall that the exact propagator for this problem is
$G(R'',u'',t;R',u') = g_0^M(R''-R',t)\left[ g_0^m(u''-u',t)-g_0^m(u''+u',t)
\right]$, with $g_0^\mu(y,t)\equiv (\mu/2\pi i \hbar)^{1/2} \allowbreak\* \exp
(i\mu y^2/2t\hbar)$, the free propagator. To a good approximation, before the
collision the wave function is given by the integral involving
$g_0^m(u''-u',t)$, after the collision by that involving $g_0^m(u''+u',t)$.
Thus to get the final wave function, one reverses \mathname{u}.

   When re-expressed in terms of $x$ and $X$, $\Psi_F = \Gamma(X(1-2\delta) +
2x\delta) \allowbreak \* \Phi(-x(1-2\delta)+2X\gamma)$, suggesting that the
final wave function has become entangled. For interactions more general than
the hard wall there will be more complicated changes in the functions, but
since the separate evolution of $u$ and $R$ follows from momentum conservation
and the general nature of the two-particle interaction, there is no getting
away from the intertwining.

   The form we take for the wave function is

$$\Psi_I = \sqrt{\NN }\exp\left(-{X^2\over 4 \Sigma^2}\right)
               \exp\left(-{x^2\over 4 \sigma^2} +ikx \right)
 \eqno(1)$$

\NI
 with both $x$ and $X$ taking values on the entire real line. ($\NN =
1/2\pi\sigma\Sigma$, and the position spreads are $\Delta X = \Sigma$ and
$\Delta x = \sigma$, both assumed real.) In principle we should use a wave
function with \mathname{x-x_0} in place of \mathname{x} above and restrict the
relative coordinate to (say) negative values (because of the hard wall).
However, the form of the propagator given above allows us to use the simpler
form, \eq{1}. That propagator says that one can look upon this scattering as
taking place on the entire line but with a second source at a reflected
position (this is the method of images applied to the path integral).  The wave
emanating from the image is the wave function for large positive times, and
this is the portion we wish to study. In particular we look at its inner
product with a test wave function (what you would have taken to be the wave
function had you not treated the wall as a quantum dynamical object) and study
its entanglement properties using the measure of [\measure\refmark].  However,
the answer to neither of these questions will change with time so that we can
study the reflected wave at whatever time is most convenient and, as in \eq{1},
that time is the time for which there is symmetry in $x$. However, since this
is now to be thought of as the reflected wave moved back to an earlier time,
the {\it entire} reflected wave must be used, i.e., variables range over the
entire line. Although this trick is available only when the method of images
can be applied, the general tenor of our results does not depend on it. You can
(also in this case) start the wave packet at some (say, negative) $x_0$ and
propagate it through the impact. The only thing to be careful of (which also
applies to our argument above) is that the incoming and outgoing waves separate
on a time scale less than that for wave packet
 spreading.

   There are now two things to check: error and decoherence. To compute
``error,"  we compare the outgoing wave to what would have been
expected had the wall not been treated dynamically. To compute decoherence we
measure the degree of entanglement as defined in [\measure\refmark]. 

 \bsheader{Error}

  We examine the overlap integral of the actual $\Psi_F$ with the wave function
that would have resulted from the idealization, $x\to -x$, namely $\Psitest
\equiv \Gamma(X)\Phi(-x) = \Gamma(R-\delta u)\Phi(-R-\gamma u)$. Using \eq{1}

$$ \eqalign{
 A  ~&  \equiv \int \Psitest^*\Psi_F   
 = \int dR du 
\Gamma^*(R-\delta u)\Phi^*(-R-\gamma u) \Gamma(R+\delta u)\Phi(R-\gamma u) \cr
    &=\NN \int dR du 
       \exp\left(-{(R-\delta u)^2\over4\Sigma^2}\right)
       \exp\left(-{(-R-\gamma u)^2\over4\sigma^2} -ik(-R-\gamma u) \right)
           \cr
    &\qquad \times  \exp\left(-{(R+\delta u)^2\over4\Sigma^2}\right)
       \exp\left(-{(R-\gamma u)^2\over4\sigma^2} +ik (R-\gamma u) \right)
 \cr }$$

\NI
 We find

$$ A^{-2} =
 \left[\gamma^2 +\delta^2 + \gamma^2 \lambda + \frac{\delta^2}\lambda \right]
 \exp\left(\frac{4k^2\lambda\sigma^2}{1+\lambda}\right)
   \qquad\hbox{~with~~} \lambda \equiv {\Sigma^2\over\sigma^2}~~
 \eqno(2)$$

\NI
 Particular experiments have their own particular wave functions; hence $A$'s
deviation from 1 varies. Here we seek the minimum deviation in the face of the
interaction with the wall. To this end, we vary $\sigma$ and $\Sigma$ so as to
maximize $A$. First we study $k=0$. $A$ now depends only on $\lambda$ (not the
sigmas separately), and varying $\lambda$, we find the optimum to be given by

$$ \lmax = \frac\delta\gamma \approx \frac mM$$

\NI
 Substituting yields $A=1$. There is {\it no} error! (N.B., $\dots$ {\it only}
for $\lambda = \lmax$.) When $k\neq0$ we maximize $A$ by optimizing $\lambda$
for given $k\sigma$. We will see that the optimum $A$ {\it always falls below
unity}, by an amount of order $\delta$. For small and large $k\sigma$ analytic
forms are

$$\matrix{
      \hbox{Small~} k\sigma \hfill
      &~~~
      & \lmax \approx \delta/\gamma \hbox{~(as before)} \hfill
      &~~~
      & 1-A \approx  2\delta k^2\sigma^2  \hfill      \cr
      \hbox{Large~} k\sigma  \hfill 
      &~~~
      & \lmax \approx \delta/2k\sigma       \hfill
      &~~~
      & 1-A \approx  2\delta k\sigma      \hfill  
                                    &~~~~&~~~~&~~~~&~~~~     \cr
  }\eqno(3)$$

\NI
 These behaviors smoothly mesh at $k\sigma \sim 1$. \eq{3} represents a
lower bound on error. The factor $\delta \approx m/M$ keeps this effect small
and is reminiscent of similar factors in measurement
theory.\upref{\yanase\refmark} It may be appropriate to think of the
confinement process as one in which the system's components are constantly
bumping up against their container, so that the small $\delta$ could pick up a
large factor related to an effective frequency of such interactions.

    How bad is the error without optimization? Writing $\Sigma^2/\sigma^2
\equiv \delta e^y$ and still assuming the error to be relatively small, one
finds $(1-A)/\delta \approx \cosh y + 2k^2\sigma^2 e^y$.  


 \bsheader{Decoherence}

  This is potentially the more damaging effect. A basis independent measure of
the degree of entanglement of the particle and wall is given in
[\measure\refmark]. The degree of entanglement is 1 minus the largest
eigenvalue of $\psi^\dagger\psi$ (or $\psi\psi^\dagger$) considered as a matrix
operator with matrix indices the arguments of $\psi$.

   Because we wish to use the variable $x$ (for the system) as if it were
unentangled, the wave function variables should be $x$ and $X$.  In terms of
these

$$ \eqalign{
 \Psi_F(x,X)
  =\sqrt{\NN} \exp\Bigl\{
    -\Omega&\left[X(1-2\delta)+2\delta x\right]^2
    -\omega\left[x(1-2\gamma)+2\gamma X\right]^2 \cr
   &+ik\left(x(1-2\gamma)+2\gamma X\right) \Bigr\}
 \cr} \eqno(4)$$

\NI
 with $\NN = (2/\pi)\sqrt{\Omega\omega}$, $\Omega \equiv 1/4\Sigma^2$, and
$\omega \equiv 1/4\sigma^2$. We can form an operator (following
[\measure\refmark]) in two ways, by integrating over either $X$ or $x$. We
choose

$$ \eqalign{
F(x',x)~&\equiv
  \int dX \Psi_F^*(X.x')\Psi_F(X,x)  \cr
  =&\sqrt{2\omega\Omega\over \pi D}\exp\left\{
     -(x^2+x^{'2}){\omega\Omega\over D} -(x-x')^2{E^2\over D}
                     +ik(1-2\gamma)(x-x') \right\}
 \cr}\eqno(5)$$


$$ \hbox{with~}\quad
 D\equiv \Omega(\gamma-\delta)^2 + 4\omega\gamma^2\;,
  \qquad\hbox{and~} \rho\equiv |(\gamma-\delta)(\Omega\delta - \omega \gamma)|
 $$

\NI
 Following [\measure\refmark], we want the largest eigenvalue of $F$ (now
thought of as the integral kernel of an operator). First note that the factor
$\exp[ ik(1-2\gamma)(x-x')]$ can be dropped because it does not affect the
eigenvalue. Next observe that $F$ is almost the same as the kernel of the
propagator for the simple harmonic oscillator. Using a standard form for this
operator,\upref{\pibook\refmark} we note the following fact. The operator

$$G(x,y) \equiv \sqrt{\beta\over \pi\sinh u}\exp
         \left[ - {\beta\over\sinh u}\left[(x^2+y^2)\cosh u -2xy\right]\right]
 $$

\NI
 has the spectrum $G_n=\exp(-u(n+1/2))$, $n=0,1,2,\dots$, irrespective of 
$\beta$.
(The connection with the harmonic oscillator is
$\beta=m\omega/2\hbar$ and $\omega t=-iu$.) 
 It is now straightforward to deduce that the spectrum of $F$ is $F_n =
\left(1-e^{-u}\right) e^{-nu}$, with $n=0,1,\dots$ and $\sinh u/2 =
\sqrt{\omega\Omega}/2\rho$. From this it follows
that the largest eigenvalue of $F$ is 

$$F_0 = 1-z^2 \;, \hbox{~with~~} z= \sqrt{\frac{w^2}4+1} -\frac
w2  \ ,
  \hbox{~~and~} w\equiv \frac{\sqrt{\omega\Omega}}\rho
 $$

\NI
 For small $w$, $F_0 \sim w$, and for large $w$, $F_0 \sim 1- 1/w^2$.

   The first issue is minimizing entanglement, that is maximizing $F_0$.
Clearly $F_0$ reaches its theoretical maximum for $w=\infty$, which requires in
turn $\Omega\delta = \omega \gamma$. Recalling the definitions of $\omega$ and
$\Omega$ this brings us to the same relation, $\Sigma^2/ \sigma^2
=\delta/\gamma$, that we found when minimizing error.\upref{\bigdeal\refmark}
It is interesting that here the entanglement is strictly zero {\it even when
the momentum, $k$, is non-zero}---if there is the special matching of wave
function spreads. In the absence of matching, the entanglement, hence the
decoherence, can be considerable, as indicated by $F_0\sim w$ for small $w$.

   It should be emphasized that this decoherence cuts down on the {\it
amplitude} of the wave function that can ultimately yield an accurate
computational result. From [\measure\refmark] we know that the maximum
amplitude available in a putative unentangled wave function $\psi(x)$ is
$\sqrt{F_0}$ and that for two such successive independent collisions it will be
the product of two such terms. If $F_0$ is not extremely close to 1, the effect
can build rapidly. Such behavior is to be contrasted with say, decay, where the
initial small deviation is in a phase, so that the effect of many independent
such deviations is only quadratic in each of them.

 \bsheader{Optimal coherence}

   The minimization of both error and entanglement have brought to light a
matching condition on the spreads of the system and apparatus,
$\Sigma^2/\sigma^2 = m/M$. This may be surprising. Based on the usual
idealization of macroscopic objects, one might have thought that there should
be no restriction on the {\it smallness} of \mathname{\Delta
X}.\upref{\notmerely\refmark} Aside from considerations of the sort in
[\timebook\refmark] (and for which $F_0=1$ provides an example of a ``special
state"), there is no reason to think that Nature would evolve into minimally
decohering states. Of course the constructor of a quantum computer may have a
strong interest in such minimizing. In any case it is of interest to consider
the possibility that the optimizing condition hold generally. In
[\destruct\refmark] it was observed that {\it all\/} pairs of objects could
satisfy the relation above if for each object, its mass, $\mu$, and its
position uncertainty, $\sigma_\mu$, were related by $\sigma_\mu^2 \sim 1/\mu$.
Possible justifications (kinds of environmental decohering) were considered in
[\destruct\refmark], but we here take the relation as a hypothesis and extend
it using dimensional analysis. Taking $\hbar=1$ and $c=1$, it is clear that
another length is needed, alternatively an energy or mass. For a confined
system the quantities that come to mind are an overall distance scale for the
system and the temperature. The former seems to me ill defined, and in
particular an attractive feature of the relation proposed is that it is not
vital to distinguish between ``system" and walls. Using then the temperature
($T$) and restoring $\hbar$, we find

$$ \sigma_\mu^2 \sim \frac{\hbar^2}{\mu k_{_B}T}
 \eqno(9)$$

\NI
 with $k_{_B}$ the Boltzmann constant. \eq{9} gives a mass-$\mu$ object a
packet size that is the geometric mean of its Compton wavelength and $\sim
0.2\,\hbox{cm}\,/T\;$kelvins. This does not seem inconsistent with experience.
Lower temperature allows larger coherent wave packets, distinguishing this
effect from others\upref{\grabert\refmark} where position fluctuations {\it
decrease} with decreasing temperature.  If the effective momentum, $k$, of the
small mass is the result of thermal fluctuations, then equipartition relates
this to temperature as well. We then have $ k^2\sigma^2 \sim (2\hbar^2k^2
/2\mu)/ k_{_B}T \sim 1$, independent of temperature. (For $k\sigma=1$, $1-A
\approx 1.2\,\delta$.) This suggests that in a heat bath, $\Delta p \sim
\hbar/\Delta x$, since $\langle p \rangle =0$.

 \bsheader{Limitations and extensions}

   We have shown that confinement need {\it not\/} force entanglement, but if
the confined objects strike the walls at finite velocity, there must be some
``error." It must be emphasized that the no-entanglement result depends not
only on a particular ratio of spreads for small and large system, but also on
the Gaussian form of the wave packet and on the form of the interaction with
the wall. In this article we have not explored the effect of relaxing these
assumptions. The minimizing of error relies on the same framework, so that one
could entertain the idea of reducing error through tailoring of the wave packet
or the walls. Based on preliminary exploration, I would say that more
complicated wave packets or walls only increase both entanglement and error.

    For application of the bounds presented here it is desirable to identify
the wall mass, \mathname{M}. Even for a steel vacuum chamber one would not
look to the mass of the entire chamber, but only the region affected by the
particle's collision, perhaps defined by the wavelength of the appropriate
phonon. For ``chambers" that are magnetic fields (etc.) one can ultimately look
to the laboratory equipment that produces these fields.

   Finally, there is our decoherence-minimizing relation, $\sigma_\mu^2 \sim
1/\mu$, or more ambitiously, $\sigma_\mu^2 \sim \hbar^2/\mu k_{_B}T$. Do
particles settle into wave packets of this size? Are two-time boundary
condition considerations (as in [\timebook\refmark]) at work? Yet another
question is the form such a relation might take for massless particles. Here
too one could ask for decoherence-minimizing scattering.

   In conclusion, we have shown that pinning a system to the table does not in
itself force entanglement with the degrees of freedom of the
container---treating the latter as a fully quantum object. Nevertheless,
subject to reasonable assumptions, that pinning will introduce ``error," in the
sense of changed outgoing wave function. Minimizing both decoherence and error
are best accomplished when a particular relation exists between the wave
function spreads of system and container. We have also computed the degree of
entanglement in situations where the minimum spread condition does not hold.

 \bsheader{Acknowledgements}
  I thank B. Gaveau, D. Mozyrsky, P. Pechukas and S. Tsonchev for helpful
discussions. This work was supported in part by the United States National
Science Foundation grant PHY 93 16681.

 \bsheader{References}

 \pritem{\chuang} Among many references, see for example, I. L. Chuang,
R. Laflamme, P. W. Shor and W. H. Zurek,
 Science {\bf 270}, 1633 (1995).

 \pritem{\timebook} L. S. Schulman, {\it Time's Arrows and Quantum
Measurement}, Cambridge Univ.\ Press, Cambridge (1997).

 \pritem{\destruct} L. S. Schulman,
 Phys.\ Lett.\ A {\bf 211}, 75 (1996). Note a misprint: ``$\epsilon$" there
should be $2m/\M$ (not $m/\M$).

 \pritem{\measure} L. S. Schulman and D. Mozyrsky, Measure of decoherence,
preprint.

 \pritem{\error} By ``decoherence" I mean loss of the primary wave function
through entanglement with other degrees of freedom, hence the inability to
interfere with portions of the wave function not so entangled. By ``error" I
mean non-entangled wave function whose value is changed from that in the
absence of the scattering.

 \pritem{\primitive} For electromagnetic or other confinement this picture will
need extension. However, the primitive underpinning of the derivation,
momentum conservation, suggests that such extension is possible.

 \pritem{\NOg} A treatment neglecting the dynamical nature of the wall would
generally omit the function $\Gamma$.


 \pritem{\yanase} M. M. Yanase,
 Phys.\ Rev.\ {\bf 123}, 666 (1961).

 \pritem{\pibook} L. S. Schulman, {\it Techniques and Applications of Path
Integration}, Wiley, New York (1981).

 \pritem{\bigdeal} Disentanglement with spread matching could have been noted
directly from $\Psi_F$, so that the surprise in being able to disentangle
despite collisions did not require [\measure\refmark]. However, measure of the
amplitude defect without matching does require those results.

 \pritem{\notmerely} There is of course Feynman's variation on the two slit
experiment (in R. P. Feynman, R. B. Leighton and M. Sands, {\it The Feynman
Lectures on Physics}, Addison-Wesley, Reading, MA (1965)), which relates
uncertainties in a macroscopic object to putative measurements of a microscopic
one. The corresponding restriction here would be that $\Delta X$ ($=\Sigma$)
not be so small that the associated $\Delta P$ destroy the small-system
interference patterns that we seek. Our optimal $\Sigma$ is far from such
values. We estimate this kinematic effect as follows. Momentum uncertainty
$\Delta P$ in the big system means uncertainty $\Delta P/M$ in the (velocity)
transformation going into the center of mass frame. For the small system this
velocity uncertainty gives a momentum uncertainty $(\Delta p)' \sim m (\Delta
P/M)$ (the prime on $(\Delta p)'$ distinguishes it from the momentum
uncertainty in the original wave function, namely $(\Delta p)_{\hbox{\romsix
usual}}\sim \hbar/\sigma$). Taking $\Delta P \sim \hbar /\Sigma$, we find
$(\Delta p)' \sim m \hbar /M\Sigma$. Using $\Sigma^2 / \sigma^2 \approx m/M$
yields $(\Delta p)'/(\Delta p)_{\hbox{\romsix usual}} \sim \sqrt{m/M}$.

 \pritem{\grabert} H. Grabert, P. Schramm and G. Ingold,
 Phys.\ Rep.\ {\bf 168}, 115 (1988). See in particular Table 2, p.\ 159.

\end